\documentclass[a4paper,11pt]{article}
\usepackage{pos}
\usepackage{xspace}
\usepackage{multirow}
\usepackage{booktabs}
\usepackage{lineno}
\usepackage{float}

\title{Assessing the QGP speed of sound in ultra-central heavy-ion collisions with ALICE}
\ShortTitle{Assessing the QGP speed of sound with ALICE}

\author*[a]{Omar Vázquez Rueda on behalf of the ALICE Collaboration}

\affiliation[a]{University of Houston,\\
Science and Research Building 1, 3507 Cullen Blvd, Houston, TX 77204}

\emailAdd{ovazquezrueda@uh.edu}

\abstract{
Ultrarelativistic heavy-ion collisions produce a state of hot, dense, strongly interacting QCD matter known as quark--gluon plasma (QGP). On an event-by-event basis, the volume of the QGP in ultra-central collisions is mostly constant, while its total entropy can vary significantly, leading to variations in the temperature of the system. Exploiting this unique feature of ultra-central collisions allows us to interpret the correlation between the mean transverse momentum $(\meanpt)$ of produced charged hadrons and the number of charged hadrons as a measure of the the speed of sound, \cs. The speed of sound, \cs, which relates to the speed at which compression waves travel in a medium (in this case the QGP), is determined by fitting the relative increase of \meanpt with respect to the relative change of the average charged-particle density\,$(\avdndeta)$ measured at midrapidity. This study reports the \meanpt of charged particles in ultra-central Pb--Pb collisions at a center-of-mass energy of 5.02 TeV per nucleon pair, using the ALICE detector. Different centrality estimators based on charged-particle multiplicity or the transverse energy of the event are employed to select ultra-central collisions. By ensuring a pseudorapidity gap between the region used to define the centrality and the region used for measurement, the influence of biases and their potential effects on the rise of the mean transverse momentum are tested. The measured $\scs$ is found to strongly depend on the type of the centrality estimator, ranging from $0.113 \pm 0.003 \, \mathrm{(stat)} \pm 0.007 \, \mathrm{(syst)}$ to $0.438 \pm 0.001 \, \mathrm{(stat)} \pm 0.019 \, \mathrm{(syst)}$ in natural units.}

\FullConference{42nd International Conference on High Energy Physics (ICHEP2024)\\
18-24 July 2024\\
Prague, Czech Republic\\}

\begin{document}

\newcommand{\PbPb}         {\mbox{Pb--Pb}\xspace}
\newcommand{\meanpt}       {\ensuremath{\langle p_{\rm T}\rangle}\xspace}
\newcommand{\pt}           {\ensuremath{p_{\rm T}}\xspace}

\newcommand{\fivenn}       {$\sqrt{s_{\mathrm{NN}}}~=~5.02$~Te\kern-.1emV\xspace}
\newcommand{\avEZDC}       {\ensuremath{\langle E_{\mathrm{ZDC}} \rangle}\xspace}
\newcommand{\avEZN}        {\ensuremath{\langle E_{\mathrm{N}} \rangle}\xspace}
\newcommand{\avEZP}        {\ensuremath{\langle E_{\mathrm{P}} \rangle}\xspace}
\newcommand{\normdndeta}   {\ensuremath{\langle\dndeta\rangle/\langle\dndeta\rangle^{0-5\%}}\xspace}
\newcommand{\cs}           {\ensuremath{c_{s}}\xspace}
\newcommand{\scs}          {\ensuremath{c_{s}^{2}}\xspace}
\newcommand{\ntracklets}   {\ensuremath{N_{\mathrm{tracklets}}}\xspace}
\newcommand{\nch}          {\ensuremath{N_{\mathrm{ch}}}\xspace}
\newcommand{\et}           {\ensuremath{E_{\mathrm{T}}}\xspace}
\newcommand{\dndeta}       {\ensuremath{\mathrm{d}N_\mathrm{ch}/\mathrm{d}\eta}\xspace}
\newcommand{\meanptref}    {\ensuremath{\langle p_{\rm T}\rangle^{0-5\%}}\xspace}
\newcommand{\nchnorm}      {\ensuremath{\langle N_{\mathrm{ch}} \rangle^{\mathrm{norm}}}\xspace}
\newcommand{\nchknee}      {\ensuremath{\langle N_{\mathrm{ch}} \rangle^{\mathrm{knee}}}\xspace}
\newcommand{\ptnorm}       {\ensuremath{\langle p_{\mathrm{T}}\rangle^{\mathrm{norm}}}\xspace}
\newcommand{\avdndeta}     {\ensuremath{\langle\dndeta\rangle}\xspace}
\newcommand{\Npart}        {\ensuremath{N_\mathrm{part}}\xspace}
\newcommand{\avNpart}      {\ensuremath{\langle N_\mathrm{part}\rangle}\xspace}
\newcommand{\normmeanpt}   {\ensuremath{\langle p_{\rm T}\rangle/\langle p_{\rm T}\rangle^{0-5\%}}\xspace}

\newcommand{\ITS}          {\rm{ITS}\xspace}
\newcommand{\ZDC}          {\rm{ZDC}\xspace}
\newcommand{\ZDCs}         {\rm{ZDCs}\xspace}
\newcommand{\ZNA}          {\rm{ZNA}\xspace}
\newcommand{\ZNC}          {\rm{ZNC}\xspace}
\newcommand{\SPD}          {\rm{SPD}\xspace}
\newcommand{\SDD}          {\rm{SDD}\xspace}
\newcommand{\SSD}          {\rm{SSD}\xspace}
\newcommand{\TPC}          {\rm{TPC}\xspace}
\newcommand{\VZERO}        {\rm{V0}\xspace}

\maketitle

\section{Introduction}

High-energy heavy ion collisions produce a quark--gluon plasma (QGP)~\cite{ALICE:2022wpn}, a state of matter in which quarks and gluons are deconfined and not bound to hadrons. On an event-by-event basis, the volume of the QGP in ultra-central collisions is mostly constant, while the charged-particle multiplicity\,$(\nch)$ can vary significantly~\cite{Gardim:2019brr,Gardim:2019xjs}. Charged-particle multiplicity variations are interpreted as fluctuations in the entropy, which is generated early in the collision primarily through interactions of the sea gluons of the colliding nuclei. Given the QGP's fixed volume, the corresponding rise in the entropy density leads to higher temperatures $(T)$, as the entropy density is approximately proportionally to $T^{3}$ for the QCD equation of state of high temperature deconfined matter~\cite{HotQCD:2014kol}. The equation of state determines how gradients in the energy density profile give rise to pressure gradients. A fundamental quantity that characterizes the expansion is the speed of sound, \cs, which is the velocity at which a compression wave travels in a medium. In a relativistic fluid, $\scs = \mathrm{d}\,P/\mathrm{d}\,\epsilon = \mathrm{d\, ln}\,T / \mathrm{d\, ln}\,s$, where $P,\epsilon$, and $s$ are the pressure, energy density, and entropy density, respectively. Assuming that the mean transverse momentum (\meanpt) and the \nch are proportional to $T$ and $s$ of the QGP~\cite{VANHOVE1982138}, respectively, the speed of sound can be determined experimentally as, $\scs = \mathrm{d\, ln}\,\meanpt / \mathrm{d\, ln}\,\nch$~\cite{Ollitrault:2007du}. This manuscript describes the analysis to extract the \scs with ALICE~\cite{ALICE:2024oox}.   


\section{Data analysis}

This study uses data from \PbPb collisions at \fivenn collected by the ALICE detector~\cite{ALICE:2008ngc} during the Run 2 data-taking period of the LHC. The relevant subdetectors for this study are the VZERO (\VZERO), the Inner Tracking System (ITS), the Time Projection Chamber (TPC), and the Zero Degree Calorimeters (\ZDC). The \VZERO provides the interaction trigger and is employed for centrality assessment. The \ITS and \TPC detectors are also employed for centrality classification, using the number of tracklets (\ntracklets) and \nch, respectively. A proxy for the event transverse energy (\et) is also used for centrality estimation, quantified as the summed transverse mass $(m_{\mathrm{T}}=\sqrt{\pt^{2}+m_{\pi}^{2}})$, where the pion mass is assumed for all particles. The \ZDC is utilized to measure the energy of the spectator nucleons, thereby providing an estimate of the average number of participating nucleons $(\langle \Npart \rangle)$. The aim of this analysis is not only to measure the speed of sound (\cs) but also to investigate dependence of the extracted \cs on the acceptance, kinematic cuts, and the actual observable used to quantify the event activity for centrality classification~\cite{Nijs:2023bzv}. To this end, different centrality estimators sensitive to the event activity in different regions of the phase--space are defined. Table~\ref{tab:centrality_definition} lists the different estimators along with the kinematic cuts on the particles to estimate centrality, as well as the \meanpt and \avdndeta. The \ntracklets-based centrality estimators include particles with transverse momentum down to $\pt \approx 0.03~\mathrm{GeV}/c$ and do not impose an upper \pt cut. In contrast, the \nch-based centrality estimators use particles with $0.15\leq \pt <50~\mathrm{GeV}/c$. This also applies to the \et-based centrality estimators. The charged particles reaching the \VZERO neither have a lower nor an upper \pt cut. 

\begin{table}[!htbp]
\caption[]{(Second column) Labels used in the figures to identify the results from each centrality estimator. (Third column) Pseudorapidity intervals for centrality estimation. (Fourth column) Pseudorapidity interval to measure the \meanpt, and \avdndeta. (Fifth column) Minimum distance between the centrality estimation region and the region to measure \meanpt, and \avdndeta. }
\centering
\begin{tabular}{cccccc}
 \toprule
 Observable & Label & Centrality estimation & $\meanpt$ and $\avdndeta$ & Minimum $|\Delta\eta|$\\
 \midrule
 \multirow{2}{*}{\nch in \TPC} & I & $|\eta|\leq0.8$ & $|\eta|\leq0.8$ & 0 \\
                               & II & $0.5\leq|\eta|<0.8$ & $|\eta|\leq0.3$ & 0.2 \\
 \midrule
 \multirow{2}{*}{\et in \TPC} & III & $|\eta|\leq0.8$ & $|\eta|\leq0.8$ & 0 \\
                              & IV & $0.5\leq|\eta|<0.8$ & $|\eta|\leq0.3$ & 0.2 \\
 \midrule
 \multirow{3}{*}{\ntracklets in \SPD} & V & $|\eta|\leq0.8$ & $|\eta|\leq0.8$ & 0 \\
                               & VI & $0.5\leq|\eta|<0.8$ & $|\eta|\leq0.3$ & 0.2 \\
                               & VII & $0.3<|\eta|<0.6$ & $|\eta|\leq0.3$ & 0 \\
                               & VIII & $0.7\leq|\eta|<1$ & $|\eta|\leq0.3$ & 0.4 \\
 \midrule
 \nch in \VZERO & IX & $-3.7<\eta<-1.7$ and $2.8<\eta<5.1$ & $|\eta|\leq0.8$ & 0.9 \\
 \bottomrule
\end{tabular}
\label{tab:centrality_definition}
\end{table}


\section{Results}

\begin{figure}[!ht]
	\centering
	\hspace{0cm}
    \includegraphics[width=0.64\textwidth]{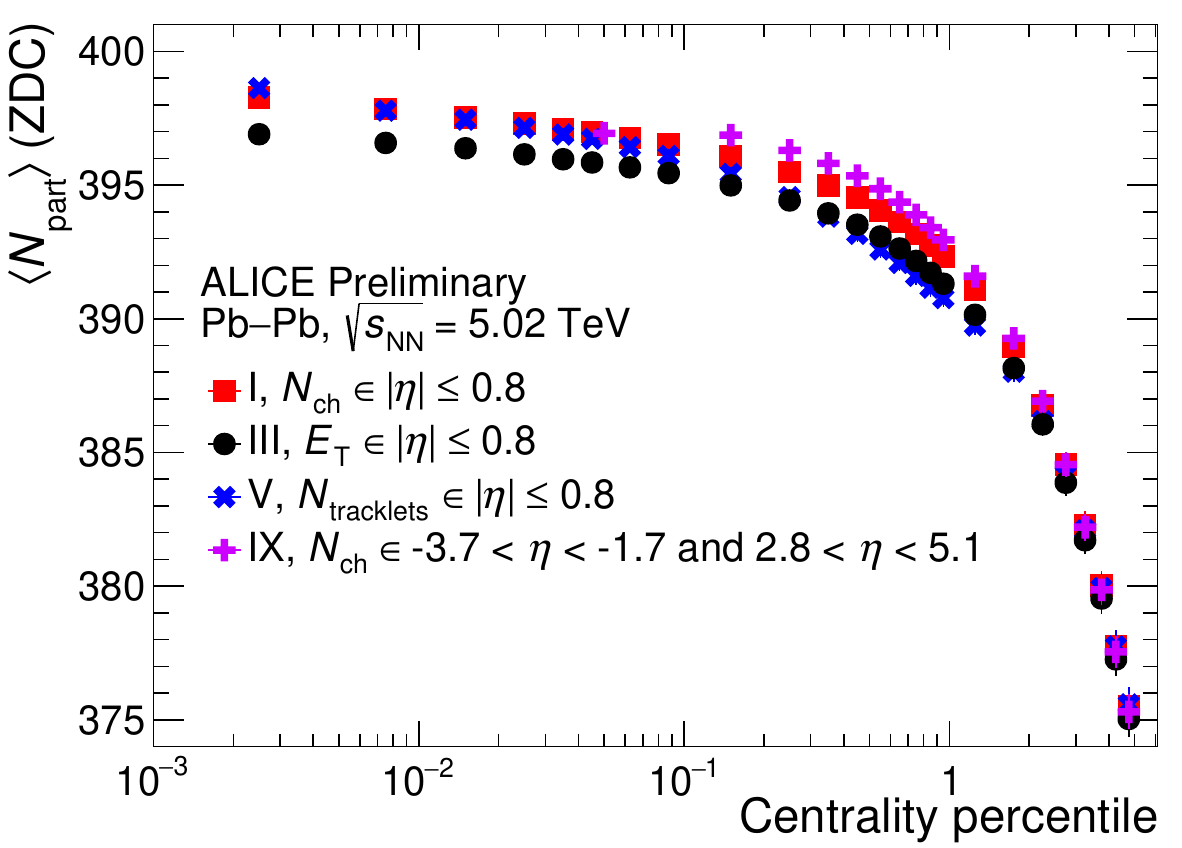}
	\hspace{0cm}
	\caption{Centrality dependent $\avNpart$. The data are shown for the I (\nch in $|\eta|\leq0.8$), III (\et in $|\eta|\leq0.8$), V (\ntracklets in $|\eta|\leq0.8$), and IX (\nch in $-3.7\leq\eta \leq-1.7+2.8\leq\eta \leq5.1$) centrality estimators. Figure taken from~\cite{ALICE:2024oox}.}
	\label{fig:Npart_vs_cent}
\end{figure}

Figure~\ref{fig:Npart_vs_cent} shows the centrality dependence of the average number of participating nucleons, \avNpart for the centrality estimators labeled I, III, V, and IX. The expression for \avNpart is given by, $\avNpart = 2\mathrm{A} - (\avEZN/\alpha_{\mathrm{N}} + \avEZP/\alpha_{\mathrm{P}}) /E_{\mathrm{A}}$, where $\mathrm{A} = 208$  is the mass number of the Pb nucleus, $E_{\mathrm{A}}=2.51~\mathrm{TeV}$ is the beam energy per nucleon, $\langle E_{\mathrm{N}} \rangle\,(\langle E_{\mathrm{P}} \rangle)$ is the energy deposited by neutrons\,(protons) in the \ZDC, and $\alpha_{\mathrm{N}}=1\,(\alpha_{\mathrm{P}}=0.70\pm0.05)$ is the neutrons\,(protons) acceptance correction~\cite{ALICE:2020bta}. The centrality-dependent \avNpart shows a common trend among the different centrality estimators, regardless of whether \nch or \et is used for event classification. The \avNpart in the $0-1\%$ centrality percentile interval exhibits a hint of a saturation, increasing by about one percent when moving from the $0.9-1\%$ to the $0-0.005\%$ centrality interval. This suggests that the volume of the QGP is mostly constant in the ultra-central collisions limit. Furthermore, the \avNpart values in the $0-0.1\%$ centrality range obtained with the \et-based centrality estimator are systematically lower than those obtained from using the \nch-based centrality estimators. This indicates different selection biases; the \et centrality estimator selects events with fewer charged particles at midrapidity than the \nch estimators for the same centrality interval.

The main observable is the centrality dependent correlation between $\normmeanpt$ and $\normdndeta$, where $\meanpt^{0-5\%}$ and $\avdndeta^{0-5\%}$ are measured in the $0-5\%$ centrality interval. Both quantities are derived from the \pt spectra in the range of $\pt=0$ to $10~\mathrm{GeV}/c$. The spectra are fully corrected for acceptance, tracking inefficiencies, and secondary particle contamination. The squared speed of sound, \scs, is extracted from a fit to the \normmeanpt versus \normdndeta correlation using the parameterization proposed in~\cite{Gardim:2019brr}.

\begin{figure}[!ht]
	\centering
	\hspace{0cm}
    \includegraphics[width=0.49\textwidth]{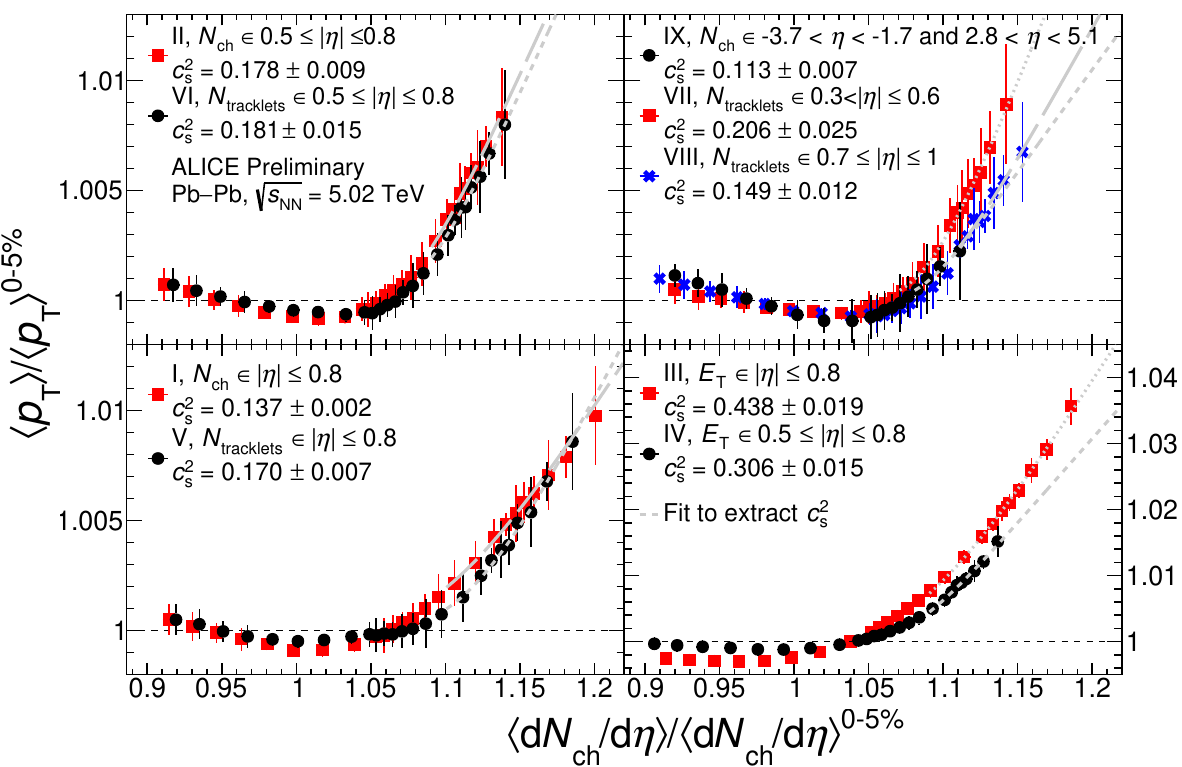}
	\hspace{0cm}
   \includegraphics[width=0.49\textwidth]{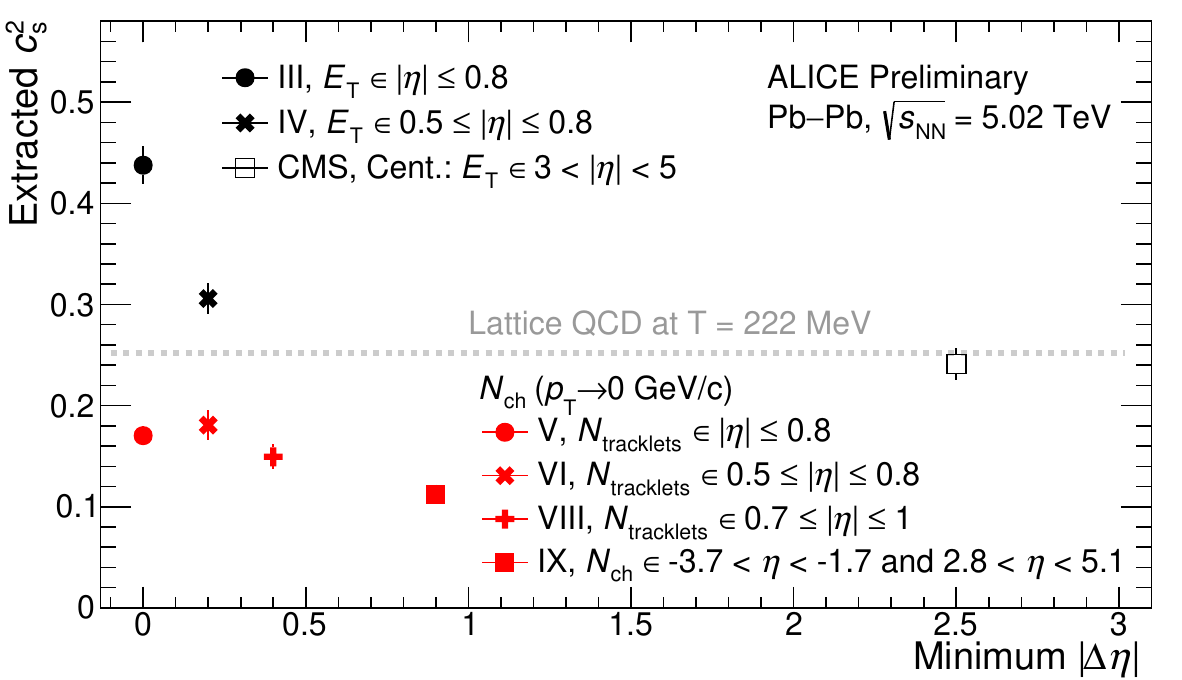}
 	\hspace{0cm}
 	\caption{(Left) Correlation between \normmeanpt and \normdndeta. The $y$-axis scale for the \et-based centrality estimators (III and IV) is to be read from the $y$-axis located to the right of the bottom right panel. The lines on top of the data correspond to fits. The uncertainty on \scs corresponds to the sum in quadrature of the statistical and systematic uncertainties. The vertical uncertainty bars in each point correspond to the sum in quadrature of the statistical and systematic uncertainty. The total uncertainty in the \normdndeta is negligible and hence not visible. (Right) Extracted \scs, as a function of the minimum $|\Delta \eta|$. The results are compared with the measured value by the CMS Collaboration~\cite{CMS:2024sgx}. The uncertainty bars around the data points correspond to the sum in quadrature of the statistical and systematic uncertainties. The Lattice QCD prediction of \scs for deconfined matter is obtained from the HotQCD collaboration~\cite{HotQCD:2014kol}. Figure taken from ~\cite{ALICE:2024oox}.}
	\label{fig:meanpT_with_fits}
\end{figure}

The left figure in Fig.~\ref{fig:meanpT_with_fits} displays the \normmeanpt versus \normdndeta for the centrality estimators listed in Tab.~\ref{tab:centrality_definition}, along with fits to the data. The top left panel shows results with $|\Delta \eta|=0.2$ using  the \SPD and the \TPC for centrality estimation. Both provide similar distributions, suggesting the yield with transverse momentum below $\pt=0.15~\mathrm{GeV}/c$ is not significant for selecting collisions with similar entropy densities. This is further confirmed by the similar \scs between the two estimators. The top right panel of Fig.~\ref{fig:meanpT_with_fits} presents results from introducing a minimum $|\Delta\eta|=0, 0.4, 0.9$. The \normmeanpt with $|\Delta\eta|=0$ rises at a steeper rate compared to when $|\Delta\eta|=0.4$, while the \normdndeta remain similar. This may be attributed to the finite width of jets, whose fragmentation products leak into the region where the \meanpt is measured. The \VZERO ensures a minimum $|\Delta\eta|=0.9$, yielding the shortest \normdndeta at midrapidity. The \normmeanpt with the \VZERO centrality estimator shows an increase but at lower rate, which is reflected in the lowest extracted \scs. The bottom left panel of Fig.~\ref{fig:meanpT_with_fits} presents results without a pseudorapidity gap and with full overlap between the pseudorapidity regions to estimate collision centrality and measure the spectra. The \normdndeta with the \TPC estimator (centrality estimator labeled I) is the highest, due to the fact that \TPC tracks are used for both centrality determination and \pt spectra, introducing a multiplicity bias; any local fluctuations (including measurement uncertainties) directly affect the \normdndeta axis. The \normmeanpt remains almost unchanged compared to when $|\Delta\eta|=0.2$ (centrality estimator labeled II). This feature ``stretches'' the distribution along the $x$-axis, resulting in a lower extracted \scs than when $|\Delta\eta|=0.2$. The bottom right panel of Fig.~\ref{fig:meanpT_with_fits} shows results obtained using the \et-based centrality estimators. The centrality estimator with full overlap between the pseudorapidity regions (centrality estimator labeled III) introduces a \pt bias that increases the \normmeanpt by up to 25\% compared to the \nch-based estimators with full overlap. Furthermore, the high \normdndeta reach suggests an additional fragmentation bias. Introducing a $|\Delta\eta|=0.2$ (centrality estimator labeled IV) reduces the \normdndeta and causes the increase in \normmeanpt to be less steep compared to when there is full overlap, although the extracted \scs is higher than the one obtained with the \nch-based estimators. This could be attributed to an interplay between the finite width of the jets and the transverse momentum bias. Figure~\ref{fig:meanpT_with_fits} shows the dependence of the extracted \scs as a function of the minimum $|\Delta\eta|$. When \et is used for centrality classification, a \pt bias is introduced, causing a rapid increase in \normmeanpt and consequently larger extracted \scs values compared to those obtained with the \nch-based centrality estimators. This can be attributed to the finite width of jets contributing particles of intermediate to high \pt to the spectra. Fig.~\ref{fig:meanpT_with_fits} also presents the \scs measured by the CMS Collaboration~\cite{CMS:2024sgx}, where the centrality determination is made in the forward pseudorapidity region $(3\leq|\eta|\leq5)$ using the top $0-5\%$ events with the highest \et. The \normmeanpt versus \normdndeta correlation is measured for $|\eta|<0.5$. The CMS Collaboration reports $\scs=0.241\pm 0.002~(\mathrm{stat})~\pm~0.016~(\mathrm{syst})$ in natural units, which is consistent with expectations from Lattice QCD~\cite{HotQCD:2014kol}. The CMS value lies between the \scs measured using the charged multiplicity and transverse energy centrality estimators from ALICE data. The CMS experimental setup has a much wider pseudorapidity gap between the centrality and observable pseudorapidity regions compared to the ALICE setup, which helps suppress short-range \meanpt--\meanpt correlations due to the finite width of jets. However, using the \et-based centrality estimator along with a wide pseudorapidity gap is still sensitive to long-range \meanpt--\meanpt correlations~\cite{Chatterjee:2017mhc}.

\section{Conclusions}

The speed of sound in ultra-central \PbPb collisions at \fivenn is extracted through a fit to the correlation between \normmeanpt and \normdndeta. The rise of the normalized \meanpt demonstrates a strong dependence on the definition of the centrality estimator. The slope of the \normmeanpt is steeper when using the \et-based centrality estimators compared to the results from the \nch-based centrality estimators. This difference is mainly attributed mainly to the effects of short- and long-range \meanpt--\meanpt correlations. These features confirm a prediction from the Trajectum hydrodynamic model~\cite{Nijs:2023bzv} and highlight the need for a reevaluation of how the speed of sound can be extracted from heavy-ion data.

\section{Acknowledgments}
The author acknowledges the support of the University of Houston for the postdoctoral fellowship. 

\bibliographystyle{utphys}   
\bibliography{bibliography}

\providecommand{\href}[2]{#2}\begingroup\raggedright\begin{thebibliography}{10}

\bibitem{ALICE:2022wpn}
{\bfseries ALICE} Collaboration, ``{The ALICE experiment -- A journey through QCD}'', \href{http://arxiv.org/abs/2211.04384}{{\ttfamily arXiv:2211.04384 [nucl-ex]}}.

\bibitem{Gardim:2019brr}
F.~G. Gardim, G.~Giacalone, and J.-Y. Ollitrault, ``{The mean transverse momentum of ultracentral heavy-ion collisions: A new probe of hydrodynamics}'', \href{http://dx.doi.org/10.1016/j.physletb.2020.135749}{{\em Phys. Lett. B} {\bfseries 809} (2020) 135749}, \href{http://arxiv.org/abs/1909.11609}{{\ttfamily arXiv:1909.11609 [nucl-th]}}.

\bibitem{Gardim:2019xjs}
F.~G. Gardim, G.~Giacalone, M.~Luzum, and J.-Y. Ollitrault, ``{Thermodynamics of hot strong-interaction matter from ultrarelativistic nuclear collisions}'', \href{http://dx.doi.org/10.1038/s41567-020-0846-4}{{\em Nature Phys.} {\bfseries 16} no.~6, (2020) 615--619}, \href{http://arxiv.org/abs/1908.09728}{{\ttfamily arXiv:1908.09728 [nucl-th]}}.

\bibitem{HotQCD:2014kol}
{\bfseries HotQCD} Collaboration, A.~Bazavov {\em et~al.}, ``{Equation of state in ( 2+1 )-flavor QCD}'', \href{http://dx.doi.org/10.1103/PhysRevD.90.094503}{{\em Phys. Rev. D} {\bfseries 90} (2014) 094503}, \href{http://arxiv.org/abs/1407.6387}{{\ttfamily arXiv:1407.6387 [hep-lat]}}.

\bibitem{VANHOVE1982138}
L.~{Van Hove}, ``Multiplicity dependence of pt spectrum as a possible signal for a phase transition in hadronic collisions'', \href{http://dx.doi.org/https://doi.org/10.1016/0370-2693(82)90617-7}{{\em Physics Letters B} {\bfseries 118} no.~1, (1982) 138--140}.

\bibitem{Ollitrault:2007du}
J.-Y. Ollitrault, ``{Relativistic hydrodynamics for heavy-ion collisions}'', \href{http://dx.doi.org/10.1088/0143-0807/29/2/010}{{\em Eur. J. Phys.} {\bfseries 29} (2008) 275--302}, \href{http://arxiv.org/abs/0708.2433}{{\ttfamily arXiv:0708.2433 [nucl-th]}}.

\bibitem{ALICE:2024oox}
{\bfseries ALICE} Collaboration, ``{Assessing the speed of sound in Pb\textendash{}Pb collisions with ALICE}'',. \url{http://cds.cern.ch/record/2904102}.

\bibitem{ALICE:2008ngc}
{\bfseries ALICE} Collaboration, K.~Aamodt {\em et~al.}, ``{The ALICE experiment at the CERN LHC}'', \href{http://dx.doi.org/10.1088/1748-0221/3/08/S08002}{{\em JINST} {\bfseries 3} (2008) S08002}.

\bibitem{Nijs:2023bzv}
G.~Nijs and W.~van~der Schee, ``{Ultracentral heavy ion collisions, transverse momentum and the equation of state}'', \href{http://dx.doi.org/10.1016/j.physletb.2024.138636}{{\em Phys. Lett. B} {\bfseries 853} (2024) 138636}, \href{http://arxiv.org/abs/2312.04623}{{\ttfamily arXiv:2312.04623 [nucl-th]}}.

\bibitem{ALICE:2020bta}
{\bfseries ALICE} Collaboration, ``{Data-driven model for the emission of spectator nucleons as a function of centrality in Pb-Pb collisions at LHC energies}'',. \url{http://cds.cern.ch/record/2712412}.

\bibitem{CMS:2024sgx}
{\bfseries CMS} Collaboration, A.~Hayrapetyan {\em et~al.}, ``{Extracting the speed of sound in quark\textendash{}gluon plasma with ultrarelativistic lead\textendash{}lead collisions at the LHC}'', \href{http://dx.doi.org/10.1088/1361-6633/ad4b9b}{{\em Rept. Prog. Phys.} {\bfseries 87} no.~7, (2024) 077801}, \href{http://arxiv.org/abs/2401.06896}{{\ttfamily arXiv:2401.06896 [nucl-ex]}}.

\bibitem{Chatterjee:2017mhc}
S.~Chatterjee and P.~Bozek, ``{Pseudorapidity profile of transverse momentum fluctuations in heavy ion collisions}'', \href{http://dx.doi.org/10.1103/PhysRevC.96.014906}{{\em Phys. Rev. C} {\bfseries 96} no.~1, (2017) 014906}, \href{http://arxiv.org/abs/1704.02777}{{\ttfamily arXiv:1704.02777 [nucl-th]}}.

\end{thebibliography}\endgroup


\end{document}